# *5f* Delocalization of Bulk FCC Americium and the (111) Surface: A FP-LAPW Electronic Structure Study

Da Gao and Asok K. Ray
Department of Physics
P. O. Box 19059
The University of Texas at Arlington
Arlington, TX 76019

**ABSTRACT**

The electronic properties of bulk fcc americium and the (111) surface have been investigated with the full-potential linearized augmented plane wave (FP-LAPW) method as implemented in the WIEN2K suite of programs. The study is carried out for the anti-ferromagnetic ground state of Am at different levels of theory: (1) scalar-relativity vs. full-relativity; (2) local-density approximation (LDA) vs. generalized-gradient approximation (GGA). Our results indicate that spin orbit coupling plays an important role in determining the electronic properties of both bulk fcc americium and the (111) surface. In general, LDA is found to give a higher total energy compared to GGA results. The spin orbit coupling shows a similar effect on the surface calculations regardless of the model, GGA *versus* LDA. The 5*f* localized-delocalized transition of americium is employed to explain our results. In addition, the quantum size effects in the surface energies and the work functions of fcc (111) americium ultra thin films (UTF) are also examined.

**INTRODUCTION**

Americium, like its nearest neighbor plutonium, occupies a pivotal position in the actinide series with regard to the behavior of *5f* electrons [1]. While the atomic volume of the actinides before Pu continuously decreases as a function of the increasing atomic number from Ac until Np, a sharp atomic volume increase from Pu to Am has been experimentally observed [2]. Such behavior has been explained by the localized 5*f* electrons in Am [3]. On the other hand, the experimentally observed 5*f* electron delocalization from Am II to Am III [4] has not been observed in a recent density-functional electronic calculation [5] with respect to the high-pressure behavior of americium. Such controversies clearly indicate that further experimental and theoretical studies are necessary to improve our understanding of americium. Recently, Griveau *et al.* [6] have studied superconductivity in Am as a function of pressure and the role of a Mott-type transition.

One effective way to probe the americium *5f* electron properties and their roles in the chemical bonding is to study their surface properties. The unusual aspects of the bonding in bulk Am are apt to be enhanced at a surface or in a thin layer of Am adsorbed on a substrate, as a result of the reduced atomic coordination of a surface atom and the narrow bandwidth of surface states. Thus, Am surfaces and thin films may also provide valuable information about the bonding in Am. However, to the best of our knowledge, *no* Am surface study exists in the literature.

Motivated by the above facts and our continuing interests in actinide surface chemistry and physics [7], this study has thus focused on the (111) surface of Am II with a fcc crystal structure, and a comparison of the properties of the 5*f* electrons properties between the fcc Am (111) surface and the fcc Am bulk in order to obtain more insight in the Am 5*f* electron behaviors. In addition, the quantum size effects in the surface energies and the work functions of the fcc Am (111) surface have been examined in detail. This paves the way for future chemisorption studies on the fcc Am (111) surface.

## COMPUTATIONAL METHOD

The computational formalism used in this study is an all-electron full-potential method with a mixed basis set of linearized-augmented-plane-wave (LAPW) and augmented-plane-wave plus local orbitals (APW+lo), with and without spin-orbit coupling (SO) [8]. In the WIEN2k code, the alternative basis set APW+lo is used inside the atomic spheres for the chemically important orbitals that are difficult to converge, whereas LAPW is used for others. Both the generalized-gradient-approximation (GGA) to density functional theory (DFT) with a gradient corrected Perdew–Berke–Ernzerhof (PBE) functional and the local density approximation (LDA) to density functional theory are used to compare the effects of GGA *vs.* LDA in the descriptions of the electronic structure properties of fcc Am [9]. For relativistic effects, core states are treated fully relativistically and for valence states, two levels of treatments are implemented: (1) a scalar relativistic scheme including the mass-velocity correction and the Darwin s-shift and (2) a fully relativistic scheme with spin-orbit coupling included in a second-variational treatment using the scalar-relativistic eigen-functions as basis. Due to severe demands on computational resources, the lattice constant was not optimized and all computations have been carried out at the experimental lattice constant 9.26 a.u. of fcc Am. A constant muffin-tin radius ($R_{mt}$) of 2.60 a.u and large plane-wave cut-off $K_{max}$ determined by $R_{mt}K_{max}$=9.0 are used for all calculations. The Brillouin zone is sampled on a uniform mesh with 104 irreducible K-points for fcc bulk americium. The (111) surface of fcc americium is modeled by periodically repeated slabs of $N$ Am layers (with one atom per layer and $N$=1-5) separated by a vacuum gap of 80 a.u. Twenty-one irreducible K-points have been used for reciprocal-space integrations in the surface calculations. For both bulk and surface calculations, the energy convergence criterion is set at 0.01 mRy. All computations have been carried out at the theoretical ground anti-ferromagnetic state [5, 10, 11], though the ground state of Am is experimentally believed to be nonmagnetic [12].

## RESULTS AND DISCUSSIONS

As shown in Fig. 1, the total energies have been calculated for both bulk fcc Am and the (111) ultra thin films up to five layers at four theoretical levels, namely, LDA-AFM-NSO, LDA-AFM-SO, GGA-AFM-NSO, and GGA-AFM-SO. The GGA is found to give a much lower total energy, about 30 Ry/atom lower, compared to the LDA calculations in both scalar relativistic (without spin-orbit coupling) and fully relativistic (with spin-orbit coupling) levels of theory. As expected, spin orbit coupling is found to lower the total energies of all films and the bulk in both LDA and GGA results. Spin orbit coupling energies have been calculated according to

$$E_{so} = E_{tot}(NSO) - E_{tot}(SO), \qquad (1)$$

and are listed in Table I. It is also observed that the total energy of films decreases with the increase in the number of layers.

The dependence of the work function and surface energy on the film thickness has also been investigated and plotted in Fig. 2 and Fig. 3. The work function $W$ is calculated from

$$W = V_0 - E_F, \qquad (2)$$

where $V_0$ is the Coulomb potential energy at the half height of the slab including the vacuum

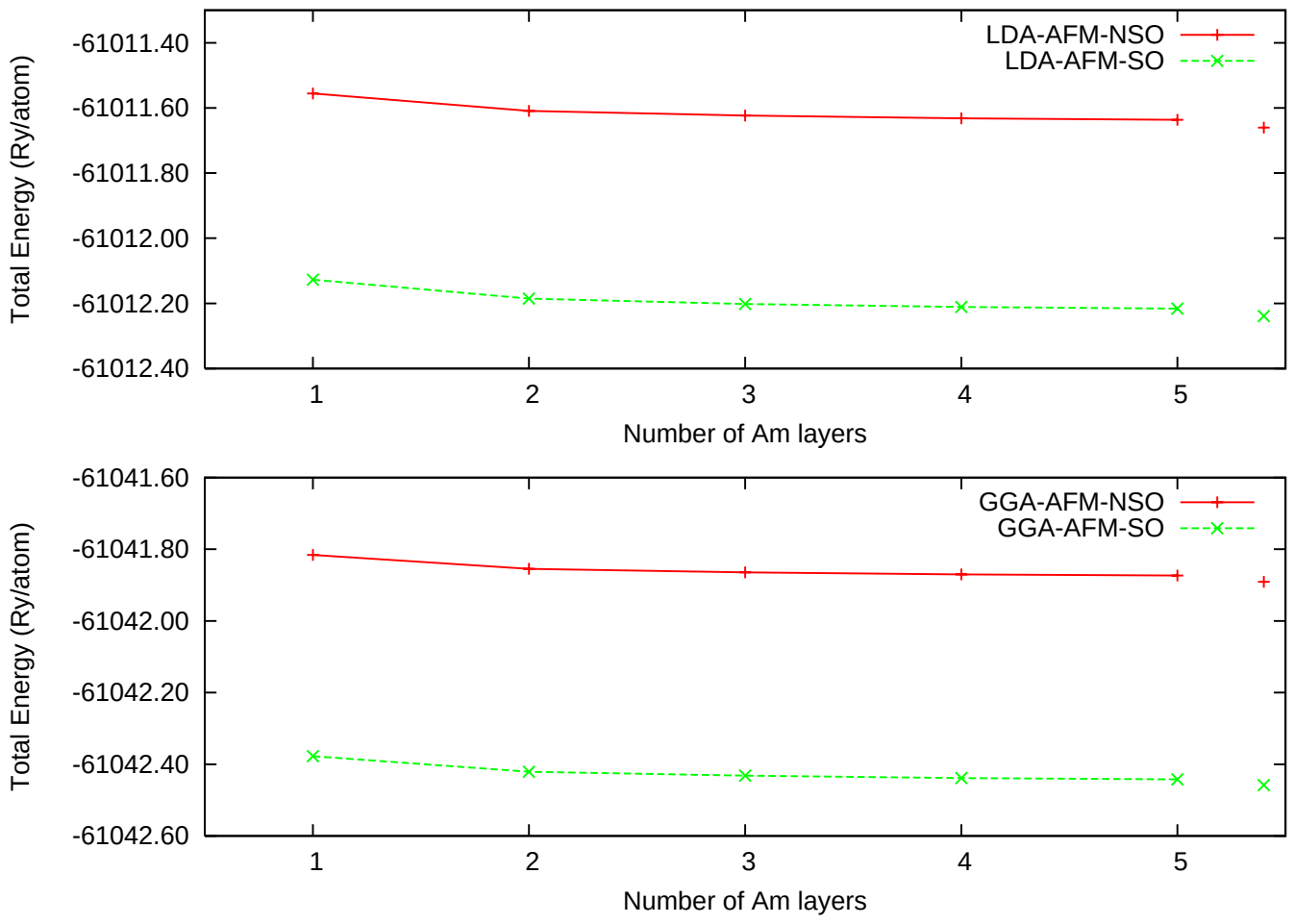

Fig.1. Total energies of Am (111) films for different layers ($N$=1-5). The corresponding total energies of Am bulk are also plotted at the farthest right of the figure.

layer and $E_F$ is the Fermi energy. The surface energy for a $N$-layer film has been estimated from

$$E_s = (1/2)[E_{tot}(N) - NE_B], \quad (3)$$

where $E_{tot}(N)$ is the total energy of a $N$-layer slab and $E_B$ is the energy of the infinite crystal. If $N$ is sufficiently large and $E_{tot}(N)$ and $E_B$ are known to infinite precision, Eq. (3) is exact.

TABLE I. Calculated work function W, surface energy $E_s$, spin magnetic moments MM, spin obit coupling energy $E_{so}$, cohesive energy $E_{coh}$ and incremental energy $E_{inc}$ of fcc Am (111) films.

| N | Theory | W (eV) | $E_s$ (J/m$^2$) | MM ($\mu_B$/atom) | $E_{so}$ (eV/atom) | $E_{coh}$ (eV/atom) | $E_{inc}$ (eV) |
|---|---|---|---|---|---|---|---|
| 1 | LDA-AFM-NSO | 3.16 | 1.05 | 7.56 | | | |
| | LDA-AFM-SO | 3.25 | 1.15 | 7.30 | 7.78 | | |
| | GGA-AFM-NSO | 2.94 | 0.75 | 7.69 | | | |
| | GGA-AFM-SO | 3.00 | 0.84 | 7.44 | 7.64 | | |
| 2 | LDA-AFM-NSO | 3.23 | 0.99 | 0.00 | | 0.73 | 1.45 |
| | LDA-AFM-SO | 3.34 | 1.09 | 0.00 | 7.84 | 0.79 | 1.58 |
| | GGA-AFM-NSO | 2.99 | 0.68 | 0.00 | | 0.53 | 1.07 |
| | GGA-AFM-SO | 3.12 | 0.78 | 0.00 | 7.70 | 0.59 | 1.18 |
| 3 | LDA-AFM-NSO | 3.20 | 1.03 | 2.53 | | 0.92 | 1.32 |
| | LDA-AFM-SO | 3.31 | 1.12 | 2.39 | 7.87 | 1.01 | 1.45 |
| | GGA-AFM-NSO | 2.95 | 0.72 | 2.59 | | 0.66 | 0.92 |
| | GGA-AFM-SO | 3.01 | 0.81 | 2.50 | 7.73 | 0.74 | 1.05 |
| 4 | LDA-AFM-NSO | 3.21 | 1.02 | 0.00 | | 1.04 | 1.38 |
| | LDA-AFM-SO | 3.36 | 1.11 | 0.00 | 7.88 | 1.14 | 1.51 |
| | GGA-AFM-NSO | 2.93 | 0.71 | 0.00 | | 0.74 | 0.99 |
| | GGA-AFM-SO | 3.04 | 0.80 | 0.00 | 7.73 | 0.84 | 1.11 |
| 5 | LDA-AFM-NSO | 3.19 | 1.03 | 1.50 | | 1.10 | 1.36 |
| | LDA-AFM-SO | 3.29 | 1.12 | 1.45 | 7.88 | 1.21 | 1.49 |
| | GGA-AFM-NSO | 2.95 | 0.72 | 1.54 | | 0.79 | 0.96 |
| | GGA-AFM-SO | 3.04 | 0.81 | 1.49 | 7.74 | 0.88 | 1.08 |

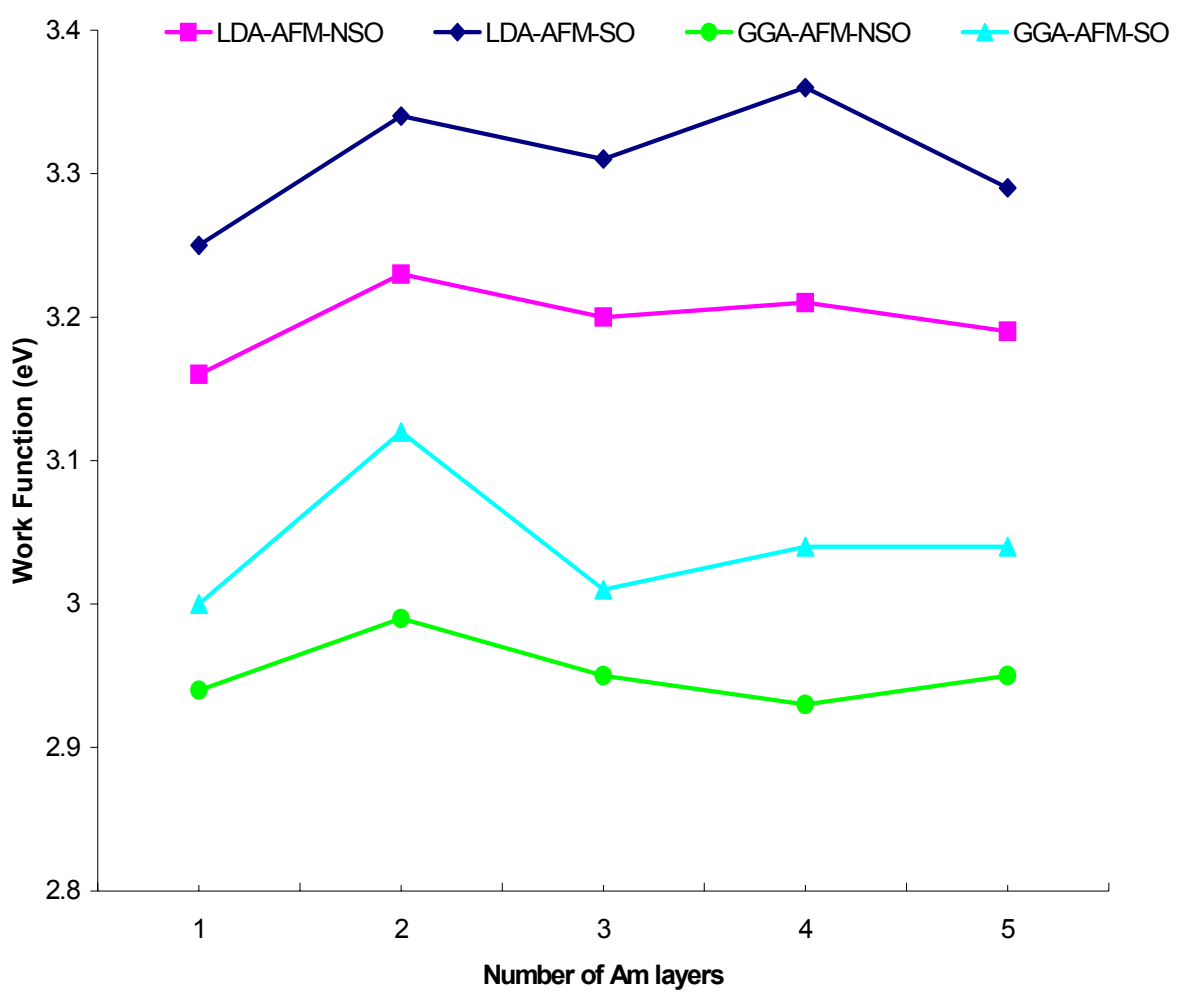

Fig. 2.Work function (eV) vs. number of Am (111) layers.

Stable and internally consistent estimates of $E_s$ and $E_B$ can be extracted from a series of values of $E_{tot}(N)$ via a linear least-squares fit to [13]

$$E_{tot}(N) = E_B N + 2E_s, \tag{4}$$

To obtain an optimal result, the fit to Eq. (4) should only be applied to films which include, at least, one bulk-like layer, *i.e.*, $N > 2$. While the work function exhibits a strong oscillation over the full range of thickness considered here, the 3 to 5-layer surface energies are relatively stable, indicating that a 3-layer film may be sufficient for future atomic and molecular adsorption studies on Am films.

The spin magnetic moments have also been calculated and the results are listed in Table I. In Fig. 4 we plotted these magnetic moments for Am (111) films as a function of the number of layers. The magnetic moments show a behavior of oscillation, which becomes smaller with the increase of the number of layers, and gradually the magnetic moments approach zero.

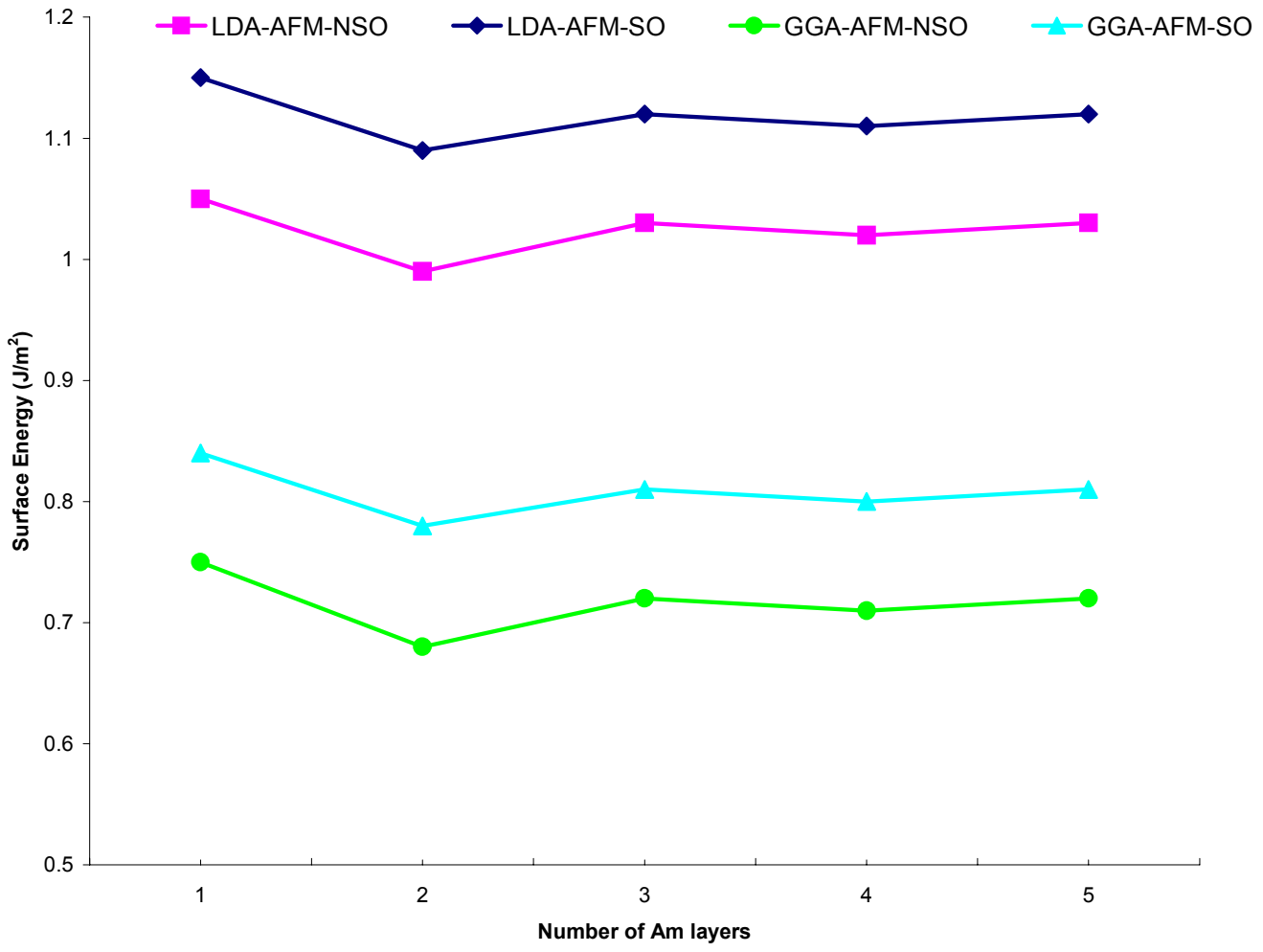

Fig.3. Surface energies for Am (111) films vs. the number of Am layers.

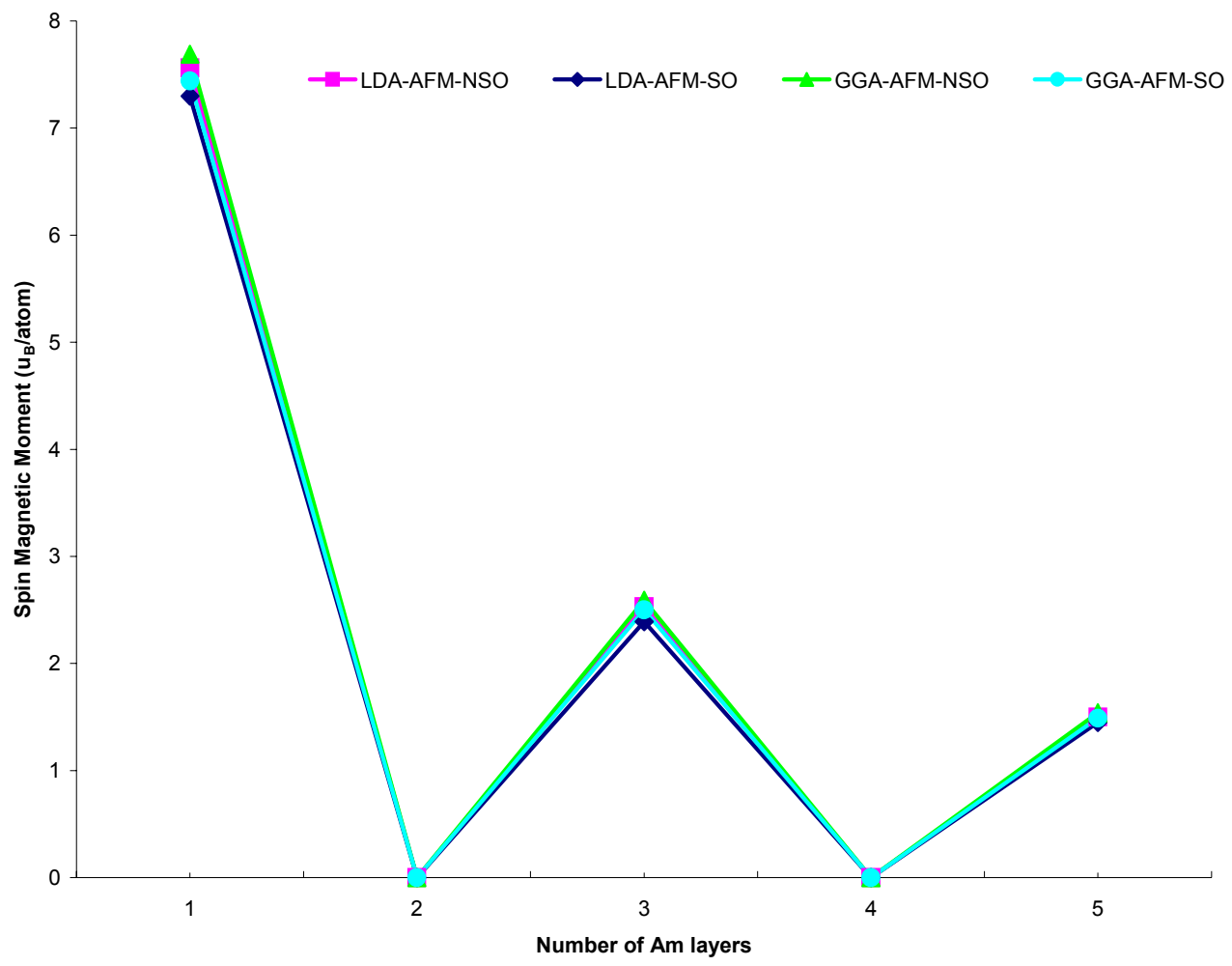

Fig. 4. Spin magnetic moments for Am (111) films as a function of the number of layers.

The cohesive energies for the Am (111) *N*-layer slabs with respect to *N* monolayers and the incremental energies of *N* layers with respect to (*N*-1) layers plus a single monolayer have been calculated at all four theoretical levels and shown in Table I as well.

In Fig. 5, we plot partial density of states of Am 5*f* electrons for bulk Am and (111) *N*-layer slabs, where *N*=1, 5 at the ground state. The 5*f* electrons in bulk Am are more localized, as indicated by a wider peak below Fermi level, compared to Am (111) films. At the GGA-AFM-SO level for the 5 layer, the first peak lies between 1 and 2eV, while the photoemission data of Naegele *et al.* [12] indicates that the 5*f* DOS should be centered over 2eV below the Fermi level. The less localized 5*f* electrons in the Am monolayer, compared to the bulk Am, indicate more 5*f* electrons participate in chemical bonding in Am (111) monolayer, and thus the relaxed lattice constants in Am (111) monolayer should be smaller than that in Am bulk. Our previous study of the Am and Pu monolayer properties [14] have shown a compression phenomena of the monolayers, which is in agreement with the present work. In addition, no apparent differences are observed between the DOS plots of LDA and GGA results.

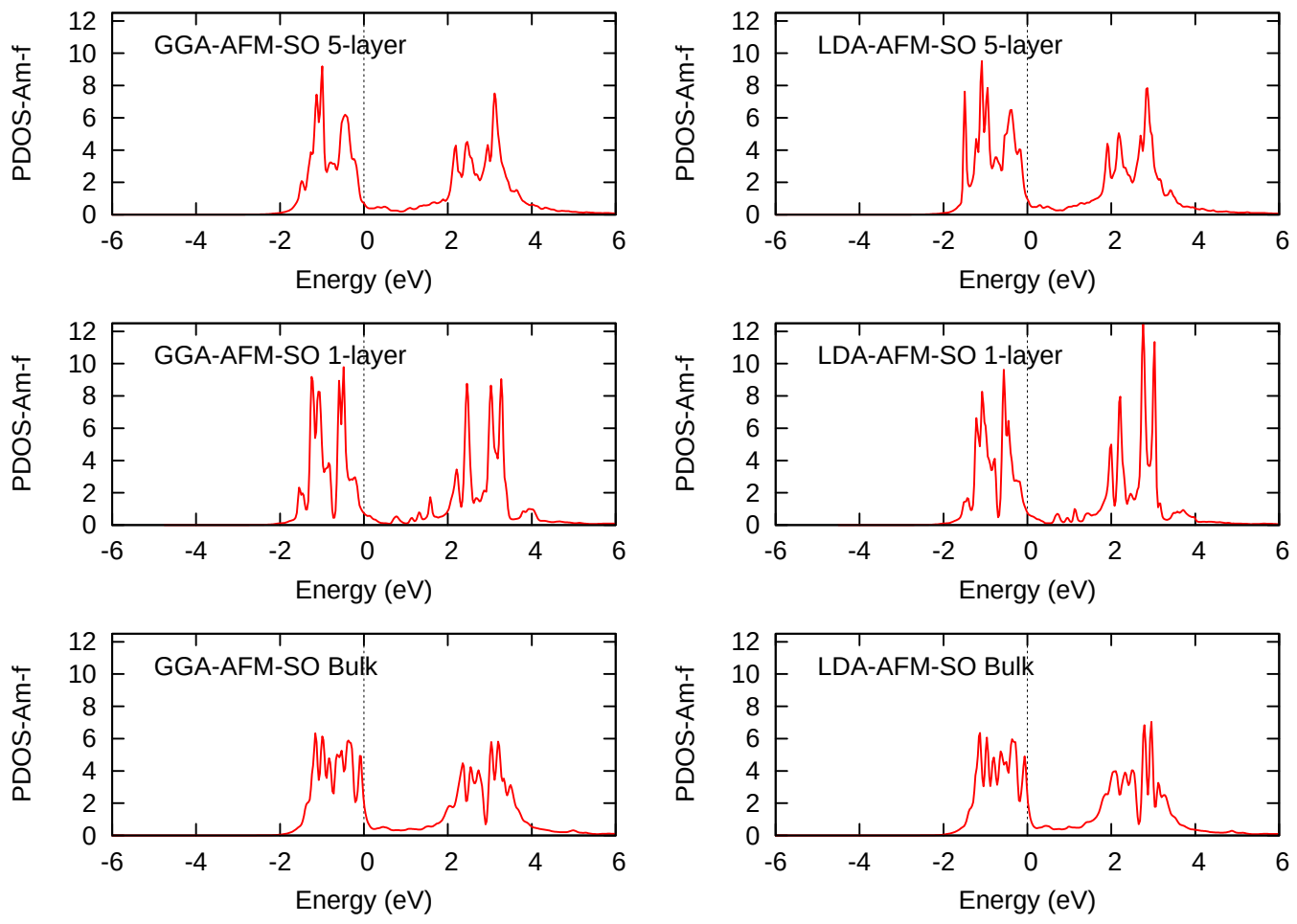

Fig.5. Density of states of 5f electrons for Am bulk and (111) *N*-layer Slabs (*N*=1, 5) at both LDA-AFM-SO and GGA-AFM-SO theoretical levels as labeled in the figure. Fermi energy is set at zero.

## CONCLUSIONS

The electronic properties of bulk fcc americium and the (111) surface have been studied with the full-potential linearized augmented plane wave (FP-LAPW) method at various levels of theory. The 5*f* electrons are found to favor delocalization in the fcc Am (111) thin films, especially in the monolayer, compared to the fcc Am bulk. The oscillatory behavior of the surface energy and work function with respect to the Am (111) film thickness suggests that a 3-layer film may provide an adequate model for estimating chemisorption energies on the fcc Am (111) surface and a thicker film would be required to realistically estimate adsorbate-induced work function shifts on the fcc Am (111) surface.

## ACKNOWLEDGMENTS

This work is supported by the Chemical Sciences, Geosciences and Biosciences Division, Office of Basic Energy Sciences, Office of Science, U. S. Department of Energy (Grant No. DE-FG02-03ER15409) and the Welch Foundation, Houston, Texas (Grant No. Y-1525).

## REFERENCES

1. J. J. Katz, G. T. Seaborg, and L. R. Morss, *The Chemistry of the Actinide Elements* (Chapman and Hall, 1986); L. R. Morss and J. Fuger, Eds. *Transuranium Elements: A Half Century A*merican Chemical Society, Washington, D. C. 1992); J. J. Katz, L. R. Morss, J. Fuger, and N. M. Edelstein, Eds. *Chemistry of the Actinide and Transactinide Elements* (Springer-Verlag, New York, in press); S. Heathman, R. G. Haire, T. Le Bihan, A. Lindbaum, K. Litfin, Y. Méresse, and H. Libotte, Phys. Rev. Lett. **85**, 2961 (2000).

2. G. H. Lander and J. Fuger, Endeavour New Series, **13**, 8, (1989).

3. A. J. Freeman and D. D. Koelling, in The *Actinides: Electronic Structure and Related Properties,* edited by A. J. Freeman and J. B. Darby, Jr. (Academic, New York, 1974) Vol. I, p. 51. B. Johansson, Phys. Rev. B. **11**, 2740 (1975).

4. A. Lindbaum, S. Heathman, K. Litfin, Y. Méresse, R. G. Haire, T. L. Bihan, and H. Libotte, Phys. Rev. B. **63**, 214101 (2001).

5. M. Pénicaud, J. Phys. Cond. Matt. **17**, 257 (2005).

6. J. -C. Griveau, J. Rebizant, G. H. Lander, and G. Kotliar, Phys. Rev. Lett. **94**, 097002 (2005).

7. A. K. Ray and J. C. Boettger, Phys. Rev. B **70**, 085418 (2004); J. C. Boettger and A. K. Ray, Int. J. Quant. Chem., **105**, 564 (2005); X. Wu and A. K. Ray, Phys. Rev. B **72**, 045115 (2005); M. N. Huda and A. K. Ray, Eur. Phys. J. B **40,** 337 (2004); Physica B **352**, 5 (2004); Eur. Phys. J. B **43**, 131 (2005); Physica B **366,** 95 (2005); Phys. Rev. B **72**, 085101 (2005); Int. J. Quant. Chem.**105,** 280 (2005); H. R. Gong and A. K. Ray, Eur. Phys. J. B, in press; H. R. Gong and A. K. Ray, submitted for publication.

8. P. Blaha, K. Schwarz, P. I. Sorantin, and S. B. Trickey, Comp. Phys. Comm. **59**, 399 (1990); M. Petersen, F. Wagner, L. Hufnagel, M. Scheffler, P. Blaha, and K. Schwarz, *ibid,* **126**, 294 (2000); K. Schwarz, P. Blaha, and G. K. H. Madsen, *ibid,* **147,** 71 (2002).

9. J. P. Perdew, K. Burke, and M. Ernzehof, Phys. Rev. Lett. **77**, 3865 (1996); J. P. Perdew and Y. Wang, Phys. Rev. B. **45**, 13244 (1992).

10. A. L. Kutepov, and S. G. Kutepova, J. Magn. Magn. Mat. **272-276,** e329 (2004).

11. D. Gao and A. K. Ray, submitted for publication.

12. J. R. Naegele, L. Manes, J. C. Spirlet, and W. Muller, Phys. Rev. Lett. **52**, 1834 (1984).

13. J. G. Gay, J. R. Smith, R. Richter, F. J. Arlinghaus, and R. H. Wagoner, J. Vac. Sci. Technol. A **2**, 931 (1984); J. C. Boettger, Phys. Rev. B **49**, 16798 (1994).

14. A. K. Ray and J. C. Boettger, Eur. Phys. J. B **27**, 429 (2002); D. Gao and A. K. Ray, Eur. Phys. J. B, in press.